\documentclass[11pt]{amsart}

\usepackage{latexsym}
\usepackage{amsfonts}
\usepackage{amsthm}
\usepackage{graphicx}
\usepackage{amssymb}
\usepackage{amsmath}
\usepackage{amssymb}
\usepackage{color}

\author[]{Dima Grigoriev}
\address{CNRS, Math\'ematiques, Universit\'e de Lille, 59655, Villeneuve d'Ascq, France}
\email{dmitry.grigoryev@math.univ-lille1.fr}

\author[]{Vladimir Shpilrain}
\address{Department of Mathematics, The City  College  of New York, New York,
NY 10031} \email{shpil@groups.sci.ccny.cuny.edu}
\thanks{Research of the second author was partially supported by
the NSF grant CNS-1117675}

\begin{document}

\title[Yao's millionaires' problem and public key encryption]{Yao's millionaires' problem and
decoy-based public key encryption by classical physics}

\begin{abstract}
We use various laws of classical physics to offer several solutions
of Yao's millionaires' problem without using any one-way functions.
We also describe several informationally secure public key
encryption protocols, i.e.,  protocols secure against passive
computationally unbounded adversary. This introduces a new paradigm
of decoy-based cryptography, as opposed to ``traditional"
complexity-based cryptography. In particular, our protocols do not
employ any one-way functions.
\end{abstract}

\maketitle

\section{Introduction}

The ``two millionaires problem" introduced by Yao in \cite{Yao} is:
Alice has a private number  $a$ and Bob has a private number $b$,
and the goal of the two parties is to solve the inequality $a  \le b
?$ without revealing the actual values of $a$ or $b$, or more
stringently, without revealing any information about $a$ or $b$
other than $a  \le b$ or $a > b$.

We note that all known solutions of this problem (including Yao's
original solution) use  one-way functions one way or another.
(Informally, a function is {\it one-way} if it is efficient to
compute but computationally infeasible to invert on ``most" inputs.)
The problem with those solutions is that it is still not known
whether one-way functions actually exist, i.e., the functions used
in the aforementioned solutions are just {\it assumed} to be
one-way.

In this paper, we offer several very simple  solutions of Yao's
millionaires' problem without using any one-way functions, but using
various laws of classical, everyday physics instead. We group our
solutions in several sections of this paper, emphasizing not only
different laws of physics employed, but also different underlying
ideas. More specifically, in Section \ref{elevator} we offer a
solution that does not even employ any particular ``laws" of
physics, but just uses a simple mechanism, an elevator in a
building. In Section \ref{vessels}, we employ the law of
communicating vessels.

In Section \ref{encryption}, we  offer  public key encryption
protocols that do not use any one-way functions and are secure
against passive computationally unbounded adversary.  Using one-way
functions is considered mandatory in the ``traditional",
complexity-based, public key cryptography. (We suggest
\cite{Menezes} as a general reference.) Security of our protocols,
on the other hand, is decoy-based (we explain below what it means)
rather than complexity-based, and this allows us to get rid of
one-way functions together with problems that accompany this
concept, including the lack of proof of the very existence of
one-way functions.

We note that in our earlier paper \cite{physics1}, we have offered
an encryption protocol with similar properties, based on principles
of electrical engineering. That paper has invited unprofessional
criticism of two kinds: (1) claims that our scheme is similar to
schemes of Kish et al. (see \cite{Kish}) because we ``use electrical
wires", like Kish does. (There are actually other similarities as
well: for example, in our paper we use letters of the English
alphabet, like Kish does); (2) suggesting incompetent
``man-in-the-middle" attacks based on misunderstanding of principles
of electrical engineering. Of course, the real purpose of both kind
of criticism was to distract attention from the main point made in
our paper \cite{physics1}, which is showing that there are secure
encryption protocols that do not employ any one-way functions, but
instead rely in their security on numerous ``decoys" of the actual
encrypted message, and this ``decoy-based" cryptography presents an
important alternative to the ``traditional", complexity-based,
cryptography. The work of Kish et al. does {\it not} use the idea of
decoy; their protocols still rely in their security on (allegedly)
one-way functions; it is just that the justification of their
functions being one-way comes from physics (e.g. they use the second
law of thermodynamics) rather than from mathematics.

An obvious advantage of decoy-based schemes is that they are secure
even against computationally unbounded (passive) adversary, which
cannot possibly be the case with complexity-based schemes. A
disadvantage is that so far, our schemes (that employ principles of
classical physics) require a physical medium, and this somewhat
limits the range of transmission. It is therefore a central question
whether or not some of our schemes based on laws of  physics can be
mimicked in the ``traditional" scenario where communicating parties
can only exchange sequences of bits with each other. Alternatively,
one can look for a physical medium (waves, or photons, or something
else) that would allow a long-range information transmission based
on similar ideas. As we point out in our Section \ref{radio}, we
believe that radio waves can be used to provide such a long-range
secure communication.

In this paper, we avoid using ``electrical wires" and try to only
use very simple physics in order to discourage ludicrous ``attacks"
and unprofessional criticism as above. Specifically, in Section
\ref{motion}, we just use classical Newton's laws of motion  for
public key encryption. This scenario, although not really practical,
provides a crystal clear illustration of our ``decoy" method.  This
particular protocol can only be used for communication over rather
short distances, but on the other hand, it relays the idea of
decoy-based encryption very clearly. It also clearly relays the
particular way to implement the general idea of decoy that we use in
the present paper, namely, combining private keys of Alice and Bob
during the transmission. We note that in our paper  \cite{physics1},
the implementation of the ``decoy" idea was different: the adversary
there faced an underdetermined system of equations, with multiple
solutions for an unknown secret number.

In Section \ref{waves}, building on the  idea of combining private
keys of the two parties, we  describe a much  more practical
protocol (using acoustic waves) that allows communication over
longer distances, although it still requires a physical medium,
which limits the range. Finally, in Sections \ref{fiber} and
\ref{radio} we speculate on how to increase the range of
transmission by using other kinds of waves.


\section{Range, private space, and private keys}
\label{private}

In the following three sections we address Yao's millionaires'
problem. We assume here that $a$ and $b$ are positive integers, such
that both are in the interval $[N_1, N_2]$  for some  $N_1, N_2 \in
{\mathbf Z_+}$.  Let $n = N_2 - N_1$. We note that this $n$ can be
made, by re-scaling, as large or as small as is convenient for a
particular approach. For example, if $n$ is too large to handle by
real-life tools, we can express $a$ as $a=a_1 \cdot m + r_a$, and
$b=b_1 \cdot m + r_b$ for some public positive integer $m<n$ and
positive integers $r_a, r_b < m$, and then compare $a_1$ to $b_1$,
etc. One can think of this as representing $a$ and $b$ in, say,
decimal form and then comparing them one digit at a time, going left
to right.

An important part of our model is the concept of a {\it private
space}. In the ``traditional" setting, where the parties communicate
over the Internet, the private space is a private computer that can,
in particular, secretly generate {\it private keys}. Without this
facility, there obviously would be no security. In our situation,
where we use real-life tools, a private space for, say, Alice is
typically a private room or other kind of container where {\it
nobody} can observe her actions. Usually, the other party (Bob) also
has a private space  where nobody can observe his actions.

On the other hand, we assume that everybody (Alice, Bob, the
eavesdroppers, if any) can observe (and measure) everything that is
going on in the ``public space", i.e., outside the union of Alice's
and Bob's private spaces.

\section{``Elevator" solution}
\label{elevator}

This is logistically the simplest solution. Suppose there is an
elevator building with at least $n = N_2 - N_1$ floors. Alice
positions herself on the floor number $a$, and Bob gets to the floor
number $b$. After that, Bob takes an elevator (Bob's private space)
going down, stopping at every floor. Alice is just watching the
elevator doors on her floor, making sure that Bob does not see her
when the elevator doors open (here is Alice's private space). If she
ever sees the elevator doors open, she knows that Bob's number is
larger. If not, then his number is smaller. Alternatively, when Bob
gets to the ground floor, he can get in touch with Alice to find out
whether she has seen the elevator doors open on her floor. That way,
both parties will end up knowing whose number is larger.

\section{``Race track" solution}
\label{race}

Here Alice and Bob run toward each other  from the opposite ends of
a race track of length $n = N_2 - N_1$. Alice maintains the speed of
$a$ m/s, and Bob maintains the speed of $b$ m/s. Whoever gets to the
midpoint of the track first, leaves a mark there and runs back,
knowing that he/she was faster, without knowing the actual speed of
the other party. Then, when the other party gets to the midpoint,
he/she will know that he/she was slower, again without knowing the
actual speed of the other party. To arrange for their private space
in this scenario, the parties have to put an inpenetrable fence
across the track at the midpoint.

The ``race track" idea can be actually implemented on a ``usual"
computer if we allow two different programs to work  with the same
file. That shared  file would be a bit string of length $n$,  with
all bits initially equal to 1. Alice provides a program that goes
over this  bit string left to right, replacing the current ``1"
symbol by ``0" at the speed of one symbol per $a$ time units. Bob
provides a similar program going over the same bit string right to
left, at the speed of one symbol per $b$ time units. When either
program replaces $\frac{n}{2}$ symbols, it replaces the current
symbol by ``X" and stops. Whose program stops first has the smaller
number.
 Note that both programs will have to use
the computer's internal clock, which is a little unusual but not
impossible.

\section{``Communicating vessels" solution}
\label{vessels}

Here we have two communicating vessels. One of them, call it $U$, is
in Alice's private space, and the other one, call it $V$, is in
Bob's private space. These vessels are connected by a horizontal
pipe attached  to their bottoms.

In the beginning the system is ``almost", but not completely, filled
with water. Then Alice starts pumping the water {\it out} of her
vessel at the speed of $a$ gallons (or whatever units) per second,
while Bob starts pumping the water {\it in} his vessel at the speed
of $b$ gallons per second. The parties are just watching whether the
level of water is decreasing or increasing. If it is decreasing,
then $a > b$; if it is increasing, then $a < b$.

\section{Encryption without one-way functions}
\label{encryption}

In this section, we describe several encryption protocols, based on
the same idea but on different laws of physics, whose security is
decoy-based rather than complexity-based. These protocols are
therefore secure even against computationally unbounded adversary.

\subsection{Using laws of motion}  \label{motion} We start with a very simple
protocol that is not practical, but on the other hand, it relays the
essence of the ``decoy" idea very clearly. Here Alice and Bob are
positioned at points A and B (respectively) of a long horizontal rod
of known mass. Alice wants to transmit to Bob her secret number
$F_a>0$.

First we describe the idea informally. Alice applies, at the point
A, a private force $F_a$ to the rod, moving it in the direction of
Bob's point B. At the same time, Bob applies, at the point B, his
private force $F_b$ in the same direction. The total force acting on
the rod therefore is $F=F_a+F_b$. This total force is public
information, i.e., anybody can measure it (by measuring the
acceleration of the rod, for example) at will. However, only Bob
knows $F_b$, so he can recover $F_a$ as $F-F_b$, while the adversary
cannot. As far as the adversary is concerned, there are too many
``decoy" values of $F_a$ because there are many ways to split public
$F$ as a sum $F=F_a+F_b$. However, for the ``decoy" to work, Alice
and Bob would have to synchronize the moment when they start
applying their private forces because if somebody goes first, the
adversary will be able to measure his/her force alone. Instead of
trying to synchronize Alice and Bob, we offer here a more
logistically feasible solution to this problem. Namely, Alice and
Bob are going to gradually (and randomly) increase their forces
until they stabilize at the values $F_a$ and $F_b$, respectively.
This strategy is also useful in foiling some of
the``man-in-the-middle" attacks, see discussion below, after the
protocol description.

Here is a more formal description of our encryption protocol.

\begin{enumerate}

\item Alice starts applying to the rod a force $F_1(t)$, which is  a random
function of time $t$,  moving  the rod in the direction of Bob's
point B.  Bob starts applying, in the same direction, a force
$F_2(t)$, which is, too, a random function of time $t$. (We do not
specify here what ``a random function of $t$" means; although this
issue deserves special attention, addressing it here would lead us
too far away from the mainstream of the paper.) When Bob starts
applying his force, he tells Alice, publicly, that he is ``in
business"; this is needed to foil a ``man-in-the-middle" attack by
impersonating Bob (see discussion below).

\item Eventually, after getting a confirmation that Bob is ``in business",
Alice stabilizes her force at $F_a$, and Bob stabilizes his force at
$F_b$. Bob detects the stabilization by observing that the rod
acceleration is not changing due to Alice's efforts for some fixed
period of time, agreed upon by both parties up front.

\item After the rod acceleration has stabilized, the force acting on
it is $F=F_a+F_b$, so Bob recovers Alice's secret $F_a$ as
$F_a=F-F_b$.

\end{enumerate}

We note once again that security of this protocol is based on the
presence of numerous ``decoy" possibilities for $F_a$, resulting
from the fact that there are many ways to split public $F$ as a sum
$F=F_a+F_b$. Different combinations of possible values of these
private keys can result in the same observable quantities in the
public space. Thus, it is impossible for the adversary to single
out, with non-negligible probability, the actual value of $F_a$
among all possible ones based on observations and measurements in
the public space. We emphasize that what makes this possible is that
the receiver (Bob) is able to influence the transmission of
information from the sender (Alice) by using his {\it private key},
as opposed to the typical scenario in complexity-based cryptography
where Bob, after having published his {\it public key}, is just
``sitting there" waiting for information from Alice to arrive.

If the adversary is {\it active} (i.e., if she is not just observing
and measuring but can interfere with the protocol itself), then she
can, of course, just mess up the transmission by applying her own
force to the rod, for example. This kind of interference cannot be
avoided in any scenario including the ``traditional" communication
over the Internet where the Internet cable can be cut. However, this
kind of interference is not so dangerous because the adversary does
not get a hold of the secret. A more dangerous kind of interference,
known as ``man-in-the-middle" attack, is where the adversary is
trying to impersonate the receiver, or sender, or both. In our
scenario, the adversary can try to impersonate Bob (the receiver),
but to compute the correct value of Alice's force $F_a$, the
adversary then would have to somehow get rid of Bob's contribution.
To prevent from being excluded from the protocol execution, Bob can
just instruct Alice not to stabilize her force at $F_a$ until he
tells her that he, too, has started to apply his force.

Thus,  our protocol is also secure against some of the
``man-in-the-middle" attacks, but we still encourage the reader to
focus on the new and important paradigm of decoy-based cryptography,
which provides security against {\it passive} computationally
unbounded adversary. Of course, this protocol can only be used for
communication over rather short distances, but on the other hand, it
relays the idea of ``decoy-based" encryption very clearly. In the
following subsections, we use the same idea (combining private keys
of Alice and Bob during the transmission) that we used in this
simple protocol, but a different physical principle  (superposition
of waves), to allow communication over longer distances.


\subsection{Using acoustic waves}  \label{waves} Now, building on the same ideas,
we are going to describe a much  more practical protocol that allows
communication over longer distances, although it still requires a
physical medium, which limits the range.

Here Alice and Bob are going to generate acoustic waves in a common
medium; one can think of an ``old-fashioned", non-digital phone
line, or some other acoustic waveguide. Alice and Bob are positioned
at points A and B (respectively) of this common medium. Alice wants
to transmit to Bob her secret number $A_1>0$, which is going to be
the amplitude of her wave. The arrangement is similar to that in our
previous subsection: Alice and Bob combine their waves (that have
the same frequency and phase) to get a wave whose amplitude $A$ is
the sum $A_1+A_2$ of the private amplitudes. Bob then recovers
Alice's secret as $A_1 = A -A_2$.

Here is a more formal description of this encryption protocol.

\begin{enumerate}

\item Alice and Bob publicly agree on the common frequency $\omega$ and
phase $\varphi$ of their waves.

\item Alice starts generating, at her point A,  a wave with frequency $\omega$ and
phase $\varphi$, while at the same time modulating the amplitude
$A(t)$ as a random function of time $t$. Bob, too, starts generating
his wave at his point B, with frequency $\omega$ and phase
$\varphi$, randomly modulating its amplitude. When Bob starts
generating his wave, he tells Alice, publicly, that he is ``in
business".

\item Eventually, after getting a confirmation that Bob is ``in business",
Alice stabilizes the amplitude of her wave at $A_1$, and Bob
stabilizes the amplitude of his wave at $A_2$.

\item After the amplitudes have stabilized, the amplitude of the
superposition of Alice's and Bob's waves is $A_1+A_2$, so Bob
recovers Alice's secret $A_1$ as $A_1 = A -A_2$.

\end{enumerate}

Security analysis here is the same as that in the previous
subsection. Again, the main point is that there are numerous
``decoy" possibilities for $A_1$, resulting from the fact that there
are many ways to split the  public amplitude $A$ as a sum
$A=A_1+A_2$. Thus, different combinations of possible values of the
private keys $A_1, A_2$ can result in the same observable quantities
in the public space, so that even a computationally unbounded
adversary cannot determine the actual secret $A_1$.

At this point we have to mention that the legend has it that the
idea of using a superposition of waves to preserve privacy of
communication (over the phone) was studied (secretly)  in Bell Labs
during World War II \cite{Bell}, as well as in the Soviet Union in
the 1950s, and possibly also in the U.K. \cite{Ellis}. However, the
idea was (allegedly) rejected because of insurmountable
technological difficulties: in the pre-digital era, the ``whole
wave" (a person's voice), and not just its amplitude or frequency,
would have to be retrieved in real time to make this idea useful in
practice.

\subsection{Using fiber optic} \label{fiber}
Communication using a fiber optic cable is one of the most widely
used ways of  transmitting information over the Internet. In this
case, pulses of light are sent through an optical fiber.  The light
forms an electromagnetic carrier wave that is modulated to carry
information. Thus, the same idea of superposition of waves
(generated by the sender and by the receiver) that we described in
Section \ref{waves} can be used in this situation as well.

\subsection{Using radio waves} \label{radio}  One can also use
other kinds of waves, provided that Alice and Bob are connected by
an appropriate waveguide (e.g. electromagnetic or optical). The
challenge now is to get rid of a waveguide in order to increase the
communication range dramatically, allowing secure communication
between, say, the planet surface and a satellite. One of the most
obvious ways to address this challenge would be using {\it radio
waves}. It would be interesting to assess technological feasibility,
in this context, of a protocol similar to that in our Section
\ref{waves}, but in any case, the theoretical idea of using
superposition of two waves to hide the secret amplitude (or
frequency) behind numerous decoys seems to be valid in this
situation as well.

\vskip .5cm

\noindent {\it Acknowledgement.} Both authors are grateful to  Max
Planck Institut f\"ur Mathematik, Bonn for its hospitality during
the work on this paper. We are also grateful to Igor Monastyrsky for
comments on physical aspects of our schemes.

\end{document}